\documentclass[10pt,journal]{IEEEtran}
\usepackage[T1]{fontenc}
\usepackage{cite}
\usepackage{flushend}
\usepackage{amsmath,graphicx}
\usepackage{amssymb}
\usepackage{algpseudocode}
\usepackage{amsfonts}
\usepackage{graphicx}
\usepackage{fancyhdr}  %
\usepackage{cases}
\usepackage{extarrows}
\usepackage{algorithm}
\usepackage{multirow,tabularx}
\usepackage{mathtools}
\usepackage{soul, color, xcolor}
\usepackage[english]{babel}
\usepackage{caption}
\usepackage{autobreak}
\usepackage{bm}
\usepackage{geometry}
\usepackage{float}  
\usepackage{stfloats}
\ifCLASSINFOpdf
\else
\fi
\begin{document}
\title{Hashing Beam Training for Near-Field Communications\vspace{0.5mm}}
\author{Yuan Xu\textsuperscript{1,2}, Li Wei\textsuperscript{3}, Chongwen Huang\textsuperscript{1,2}, Chen Zhu\textsuperscript{4}, Zhaohui Yang\textsuperscript{1}, Jun Yang\textsuperscript{5}, Jiguang He\textsuperscript{6}, 
\\Zhaoyang Zhang\textsuperscript{1} and M\'{e}rouane~Debbah\textsuperscript{7,8},~\IEEEmembership{Fellow,~IEEE}
\\\textsuperscript{1} College of Information Science and Electronic Engineering, Zhejiang University, 310027, Hangzhou, China
\\\textsuperscript{2} State Key Laboratory of Integrated Service Networks, Xidian University, 710071, Xi'an, China
\\\textsuperscript{3} School of Electrical and Electronics Engineering, Nanyang Technological University, Singapore
\\\textsuperscript{4} College of Engineering, Zhejiang University, 310015, Hangzhou, China
\\\textsuperscript{5} Wireless Product R\&D Institute, ZTE Corporation, China
\\\textsuperscript{6} Technology Innovation Institute, 9639 Masdar City, Abu Dhabi, UAE
\\\textsuperscript{7} KU 6G Research Center, Khalifa University of Science and Technology, P O Box 127788, Abu Dhabi, UAE
\\\textsuperscript{8} CentraleSupelec, University Paris-Saclay, 91192 Gif-sur-Yvette, France
\vspace{-5mm}
\thanks{
The work was supported by the China National Key R\&D Program under Grant 2021YFA1000500 and 2023YFB2904800, National Natural Science Foundation of China under Grant 62331023, 62101492, 62394292, 62231009 and U20A20158, Zhejiang Provincial Natural Science Foundation of China under Grant LR22F010002, Zhejiang Provincial Science and Technology Plan Project under Grant 2024C01033, and Zhejiang University Global Partnership Fund.
}
}

\maketitle
\thispagestyle{empty}
\pagestyle{empty}
\begin{abstract}
In this paper, we investigate the millimeter-wave (mmWave) near-field beam training problem to find the correct beam direction. In order to address the high complexity and low identification accuracy of existing beam training techniques, we propose an efficient hashing multi-arm beam (HMB) training scheme for the near-field scenario. Specifically, we first design a set of sparse bases based on the polar domain sparsity of the near-field channel. Then, the random hash functions are chosen to construct the near-field multi-arm beam training codebook. Each multi-arm beam codeword is scanned in a time slot until all the predefined codewords are traversed. Finally, the soft decision and voting methods are applied to distinguish the signal from different base stations and obtain correctly aligned beams. Simulation results show that our proposed near-field HMB training method can reduce the beam training overhead to the logarithmic level, and achieve 96.4\% identification accuracy of exhaustive beam training. Moreover, we also verify applicability under the far-field scenario.

\end{abstract}
	
\vspace{-2mm}
\section{Introduction}
\vspace{2.5mm}
Millimeter-wave (mmWave) has become a highly sought-after next-generation mobile communication technology due to its large bandwidth, high frequency, and its ability to suppress multi-path effects and clutter echo interference\cite{Khan5876482,Jameel8594703,Yang8613274}. However, the high frequency makes the signals more susceptible to penetration loss and path loss, resulting in reduced communication coverage\cite{Kutty7342886,Niu07228,Wei7000981}. An effective approach is to increase beamforming gain using large-scale antenna arrays\cite{Sefun8882285,Heath7400949}. Fortunately, this is feasible owing to the development of massive multiple-input multiple-output (MIMO) technology\cite{Katayama5779470,Hashemi8486279,Yngvesson20062}. 
\par
In a MIMO setting, the beamforming technique allows signals to be transmitted in the form of directional beams, resulting in higher spatial resolution and beam directivity. This technique requires precise knowledge of the angle of arrival (AoA) and angle of departure (AoD) of the mmWave propagation channel, which is usually acquired during the beam training stage prior to the data transmission\cite{Choi7786130,Aviles7750625,Junyi5262295,You9129778}. Among the existing training methods, the exhaustive training method requires traversing the entire beam space\cite{Junyi5262295}, which is the most accurate but has significant delay and training overhead\cite{Wang9771330}. Hierarchical training, on the other hand, consists of multiple stages. In each stage, the beam space is divided into two halves until the resolution requirement is met\cite{Hur6600706}. The downside is that there is a serious inherent error propagation problem \cite{Noh7417848}. In addition, the equal interval multi-arm beam (EIMB) training method employs a pre-determined sequence of multiple beams, and the optimal beam direction is obtained through ensemble operations after multiple rounds\cite{You9129778}. However, its fixed beam composition leads to fixed leakage interference, which limits the identification accuracy of the beam training to a certain extent.
\par 
\par
Specifically, we first construct the near-field single-beam training codebook, which maintains the interference between the training beams as small as possible. Further, for each BS, we use hash functions and jointly design the antenna responses to construct HMB codebook. At each time slot, each BS selects one multi-arm beam codeword to transmit the signal, and users record their received signals until all the predefined codewords in the HMB codebook have been traversed. Finally, soft decisions and voting based on the received signal power are applied to obtain the aligned beam. Simulation results show that our proposed near-field HMB training method can significantly improve the identification accuracy of near-field beam training to 96.4\% of the exhaustive beam training method while reducing the training overhead to the logarithmic level. Further, we validate its applicability under the far-field scenario.
\par
The rest of the paper is organized as follows. Section II introduces the channel and signal models of the interest scenario; Section III details the generation method of the training codebook, the working principle of the decision, and voting mechanisms. Section IV provides the numerical results of the proposed beam training technique. Finally, Section V is the conclusion of the paper.

\vspace{-1.5mm}
\section{System Model}
\begin{figure}[t]
	\begin{center}
			\centerline{\includegraphics[width=0.42\textwidth]{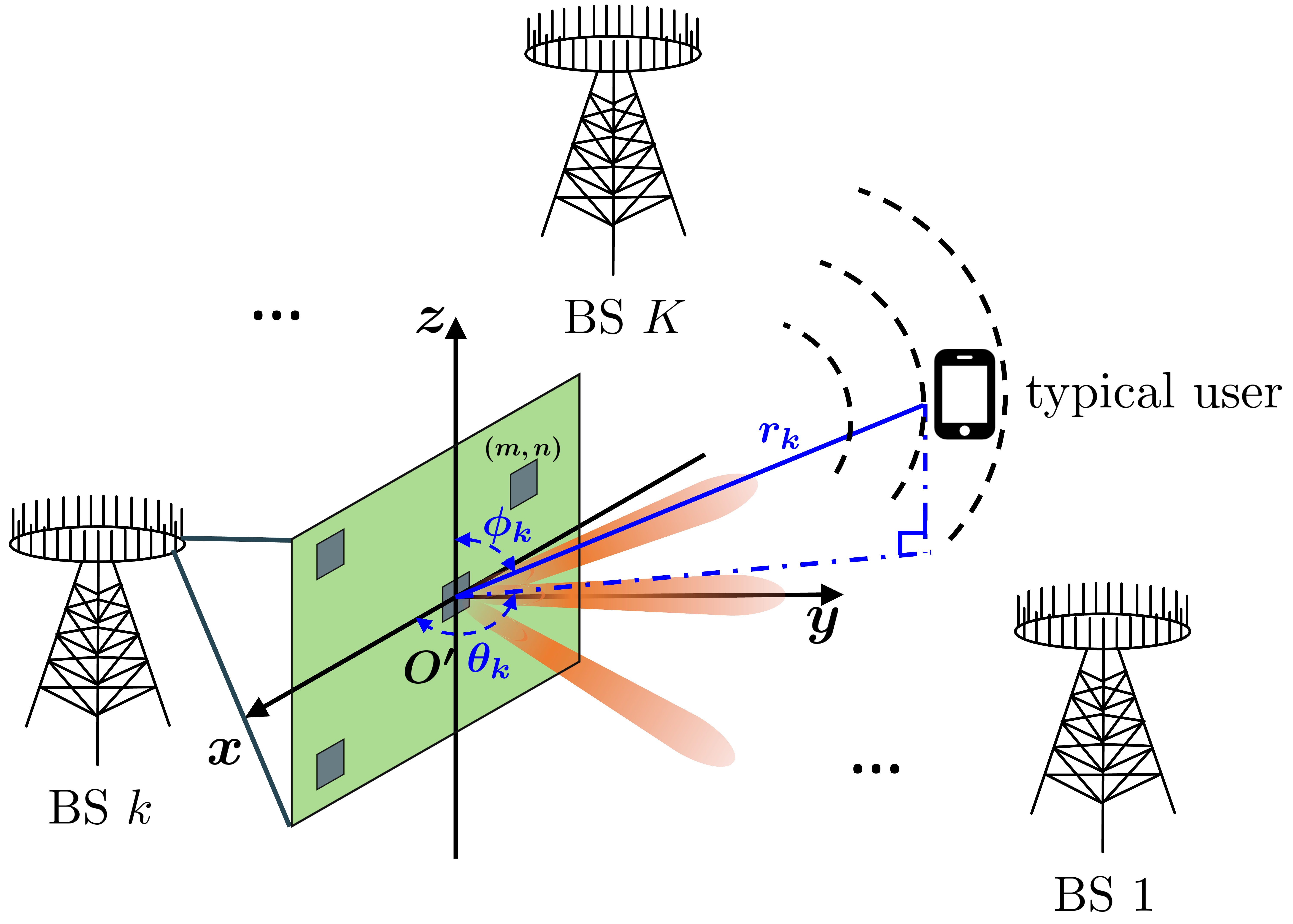}}  \vspace{-2.5mm}
			\caption{Downlink mmWave communication scenario with $K$ BSs, and a typical user. }
			\label{fig:scene} \vspace{-9mm}
		\end{center}
\end{figure}

\vspace{2.mm}
As shown in Fig.~\ref{fig:scene}, we consider a downlink mmWave communication scenario where $K$ BSs and $U$ users are distributed in the 3D space. The BS is deployed in the $xz$-plane, which employs a hybrid precoding architecture that equips $V$ radio-frequency (RF) chains and an $M\times N$ antenna uniform planar array (UPA), where $V\ll MN$. Each user device is equipped with a single antenna. The central wavelength, horizontal/vertical antenna spacing, and operating frequency are $\lambda_c$, $d_x$/$d_z$, and $f_c$, respectively. The coordinate of the $(m,n)$-th antenna element of the $k$-th BS is $(x_{k,n},y_k, z_{k,m})$, where $x_{k,n}=r_k\cos\theta_k\sin\phi_k+nd_x $, $y_k=r_k\sin\theta_k\sin\phi_k$, $z_{k,m}=r_k\cos\phi_k+md_z $, $n=1-\frac{N+1}{2},...,N-\frac{N+1}{2}$, $m=1-\frac{M+1}{2},...,M-\frac{M+1}{2}$. $r_k$, $\theta_k$, and $\phi_k$ denote the distance, azimuth angle, and elevation angle from the original $O$ to the $k$-th BS. For beam training, $K$ BSs transmit training symbols in several directions while the users listen to the channel in all directions using a quasi-omnidirectional beam. Since the alignment for each user requires only its received multi-BS superimposed signal power, we can perform downlink beam training for different users simultaneously without interference. For simplicity, we only discuss a typical user located at the $O$ position. 
\par
Let $\mathbf{h}_k\in \mathbb{C}^{MN\times 1}$ denote the channel from the BS $k$ to the user, each BS transmits the same symbol $x$ with power $P_0$, thus the received signal $y$ at the user can be expressed by\vspace{-2mm}
\begin{equation}
	y=\sum\limits_{k=1}^K\mathbf{h}_k^H\mathbf{F}_{RF_k}\mathbf{f}_{BB_k}x+n,\vspace{-1mm}
\end{equation}
where $\mathbf{f}_{BB_k}\in \mathbb{C}^{V\times1}$, $\mathbf{F}_{RF_k}\in \mathbb{C}^{MN\times V}$ denote digital precoder and analog beamformer at the $k$-th BS, $n\sim\mathcal{CN}(0,\sigma^2)$ denotes Gaussian additive white noise.
\par
We focus on the radiated near-field region. Based on the spherical wave assumption, the near-field channel can be represented as\vspace{-4mm}
\begin{equation}\label{eq:channel model1}
	\mathbf{h}_k=\sqrt{MN}\sum\limits_{l=1}^L\beta_{k,l} e^{-j\psi_{k,l}} \mathbf{g}_{k,l},\vspace{-1mm}
\end{equation}
where $\beta_{k,l}$ and $\psi_{k,l}$ denote the complex path gain and phase shift of the $l$-th path from the typical user to the $k$-th BS. In the line of sight (LoS) link dominated channel, we have $\beta_{k,1}\gg\beta_{k,l}, l\neq 1 $, thus $ \mathbf{h}_k\approx\sqrt{MN}\beta_{k,1} e^{-j\frac{2\pi}{\lambda_c}r_k}\mathbf{g}_{k,1}$, where $\beta_{k,1}=\frac{\sqrt{\rho_0}}{r_k}$ with $\rho_0$ denoting the reference channel power gain at a distance of $1$ m,
\begin{equation}
\begin{split}\label{eq:a}
    \mathbf{g}_{k,1}=&\frac{1}{\sqrt{MN}}[e^{j\frac{2\pi}{\lambda_c}(D(-\frac{M-1}{2},-\frac{N-1}{2})-r_k)},...,\\
    &e^{j\frac{2\pi}{\lambda_c}(D(m,n)-r_k)},...,e^{j\frac{2\pi}{\lambda_c}(D(\frac{M-1}{2},\frac{N-1}{2})-r_k)}]^T,
\end{split}
\end{equation}
where $D(m,n)$ denotes the distance between the $(m,n)$-th antenna of the $k$-th BS and the user. Assuming in the near-field region, $d_x/r_k\ll 1$ and $d_z/r_k\ll 1$, we can approximate $D(m,n)$ by a second-order Taylor expansion\cite{Liu10220205} as
\begin{align}\label{eq:dist_approx}
		&\quad D(m,n)\nonumber\\
  =&((r_k\cos\theta_k\sin\phi_k+nd_x)^2+(r_k\sin\theta_k\sin\phi_k)^2\vspace{-2mm}\nonumber\\
  &+(r_k\cos\phi_k+md_z)^2)^{\frac{1}{2}}\nonumber\\=&(r_k^2+n^2d_x^2+2r_knd_x\cos\theta_k\sin\phi_k\nonumber\\&+m^2d_z^2+2r_kmd_z\cos\phi_k)^{\frac{1}{2}}\nonumber\\
		\approx&  r_k+nd_x\cos\theta_k\sin\phi_k+\frac{n^2d_x^2(1-\cos^2\theta_k\sin^2\phi_k)}{2r_k}\nonumber\\
  &+md_z\cos\phi_k+\frac{m^2d_z^2\sin^2\phi_k}{2r_k}.\vspace{-2mm}
\end{align}
\par
Thus the phase of $\mathbf{g}_{k,1}$ can be divided into two parts related only to $m$ and $n$, i.e., $j\frac{2\pi}{\lambda_c}(nd_x\cos\theta_k\sin\phi_k+\frac{n^2d_x^2(1-\cos^2\theta_k\sin^2\phi_k)}{2r_k})$ and $j\frac{2\pi}{\lambda_c}(md_z\cos\phi_k+\frac{m^2d_z^2\sin^2\phi_k}{2r_k})$. Hence we obtain\vspace{-3mm}
\begin{subequations}\label{eq:channel model2}
	\begin{equation}
		\mathbf{g}_{k,1}=\mathbf{v}_x(\theta,\phi,r)\otimes\mathbf{v}_z(\phi,r),
	\end{equation}
	\begin{equation}
		[\mathbf{v}_x(\theta,\phi,r)]_n\!\!=\!e^{\!-\!j\frac{2\pi}{\lambda_c}\!(nd_x\!\cos\!\theta_k\!\sin\!\phi_k+\frac{n^2\!d_x^2\!(1-\cos^2\!\theta_k\!\sin^2\!\phi_k)\!}{2r_k})},\vspace{-2mm}
	\end{equation}   
	\begin{equation}
		[\mathbf{v}_z(\phi,r)]_m=e^{-j\frac{2\pi}{\lambda_c}(md_z\cos\phi_k+\frac{m^2d_z^2\sin^2\phi_k}{2r_k})}.\vspace{-2mm}
	\end{equation} 
\end{subequations}
where $[\cdot]_n$ denotes the $n$-th element of this vector.
\par
To find information about the user's location relative to each BS, we can design the beam-aligned observation vector $\mathbf{a}=\mathbf{F}_{RF_k}\mathbf{f}_{BB_k}$, which can be simply obtained as $\mathbf{a}=\mathbf{g}_{k,1}$.
\vspace{-1mm}
\section{Near-Field Codebook Generation}
\vspace{1.95mm}
In this section, we focus on solving the mismatch between the existing far-field codebook and the near-field. We first design a single-beam training codebook applicable to the near field, then generate a multi-arm beam training codebook using the hashing method.
\vspace{-2mm}
\subsection{Near-Field Single-Beam Codebook}
\vspace{2.4mm}
Since the scattering paths are limited and the far-field steering vectors are only angle-dependent, the channel sparsity of the far-field in the angle domain can usually be obtained by discrete Fourier transform. However, the near-field channel model in (\ref{eq:channel model2}) shows that channel $\mathbf{h}_k$ is nonlinear with respect to the antenna subscripts $m$, $n$, instead of being a discrete Fourier vector, it can be described jointly by several far-field Fourier vectors. This implies that there is severe interference between the angle-discretized training beams, and applications in the near field will suffer a significant performance loss. 
\par
Although the angular domain sparsity in the near-field region no longer exists, the number of paths is still finite. Therefore, the near-field channel is also compressible. Since channel $\mathbf{h}_k$ is determined by both the distance and angle, an intuitive idea is to reconstruct a set of orthogonal bases $\boldsymbol{a}(\theta_s,\phi_s,r_s)$ that satisfy sparsity for the near field by devising an angle- and distance-sampling approach, \vspace{-2mm}
\begin{equation}
\begin{split}\label{eq:a}
    \boldsymbol{a}(\theta_s,\phi_s,&r_s)=\frac{1}{\sqrt{MN}}[e^{j\frac{2\pi}{\lambda_c}(D^s(-\frac{M-1}{2},-\frac{N-1}{2})-r_s)},...,\\
    &e^{j\frac{2\pi}{\lambda_c}(D^s(m,n)-r_s)},...,e^{j\frac{2\pi}{\lambda_c}(D^s(\frac{M-1}{2},\frac{N-1}{2})-r_s)}],
\end{split}\vspace{-2mm}
\end{equation}
where $D^s(m,n)$ denotes the distance from the $(m,n)$-th antenna of the BS to the sampling point. 
\par
The distance and angle sampling principle is making $\eta$ as small as possible,\vspace{-2mm}
\begin{equation}
	\eta\triangleq\max\limits_{p\neq q} f(\theta_p,\theta_q,\phi_p,\phi_q,r_{p},r_{q}),\vspace{-2mm}
\end{equation}
where $f(\theta_p,\theta_q,\phi_p,\phi_q,r_{p},r_{q})=|\boldsymbol{a}(\theta_p,\phi_p,r_p)\boldsymbol{a}(\theta_q,\phi_q,r_q)^H| $ denotes the projection between the two sampled near-field steering vectors. To obtain a closed-form expression for the projection, we make an approximation similar to (\ref{eq:dist_approx}). Thus,\vspace{-1mm}
\begin{subequations}
\begin{equation}\label{eq:f}
    \begin{split}
        &f(\theta_p,\theta_q,\phi_p,\phi_q,r_{p},r_{q})=\frac{1}{MN}\\
        &|\sum\limits_{n=-\frac{N-1}{2}}^{\frac{N-1}{2}}\!\!\!\!\!\!\!e^{j\frac{2\pi}{\lambda_c}g_n(\theta_p,\theta_q,\phi_p,\phi_q,r_{p},r_{q})}\!\!\!\!\!\!\!\sum\limits_{m=-\frac{M-1}{2}}^{\frac{M-1}{2}}\!\!\!\!\!\!\!e^{j\frac{2\pi}{\lambda_c}g_m(\phi_p,\phi_q,r_{p},r_{q})}|,
    \end{split}
    \end{equation}\vspace{-1mm}
\begin{equation}\label{eq:g_m}
\begin{split}
    g_m(\phi_p,\phi_q,&r_{p},r_{q})=-md_z(\cos\phi_p-\cos\phi_q)\\&+\frac{m^2d_z^2\sin^2\phi_p}{2r_p}-\frac{m^2d_z^2\sin^2\phi_q}{2r_q},
\end{split}
    \end{equation}\vspace{-2mm}
    \begin{equation}\label{eq:g_n}
    \begin{split}
        &g_n\!(\theta_p,\!\theta_q,\!\phi_p,\!\phi_q,\!r_{p},\!r_{q})\!=\!-nd_x\!(\cos\!\theta_p\!\sin\!\phi_p\!-\!\cos\!\theta_q\!\sin\!\phi_q\!)\\
        &\!+\!\frac{n^2d_x^2(1\!-\!\cos^2\!\theta_p\sin^2\!\phi_p)}{2r_p}\!-\!\frac{n^2d_x^2(1\!-\!\cos^2\!\theta_q\sin^2\!\phi_q)\!}{2r_q}.
    \end{split}
    \end{equation}	\vspace{-1mm}
\end{subequations}
\par 
Based on (\ref{eq:f}), we can design the angular and distance sampling method as follows,\vspace{-2mm}
\begin{equation}
    \cos\phi_s=\frac{2s-M-1}{M}, \quad s=1,...,M,\vspace{-3mm}
\end{equation}
\begin{equation}\label{eq:dist_sample}
|\frac{\sin^2\phi_p}{r_p}-\frac{\sin^2\phi_q}{r_q}|\geq\frac{2\lambda_c\zeta_\Delta^2}{M^2d_z^2},\vspace{-1mm}
\end{equation}
where $\Delta$ is the projection threshold, and $|\frac{C(\zeta_\Delta)+jS(\zeta_\Delta)}{\zeta_\Delta}|=\Delta$.
Consequently, we take the observation vector $\mathbf{a}$ to be the set of orthogonal bases $\boldsymbol{a}(\theta_s,\phi_s,r_s)$, and the near-field single-beam codebook $\mathbf{C} $ can be expressed as\vspace{-1mm}
\begin{equation}
\mathbf{C}=[\boldsymbol{a}(\theta_1,\phi_1,r_1);...;\boldsymbol{a}(\theta_S,\phi_S,r_S)].\vspace{-1mm}
\end{equation}
It is worth noting that the above codebook construction method can be also applied to the far-field region when the sampling distance $r_s$ tends to infinity, which can be validated by subsequent simulations.
\vspace{-2.7mm}
\subsection{Hashing Multi-Arm Beam Codebook Generation}
\vspace{2.6mm}
To further reduce the time slot of beam training, and at the same time ensure identification accuracy, we now discuss how to generate multi-arm beams that point in multiple directions simultaneously. First, with reference to the structure of the sparse Fourier transform, we consider selecting codewords from the single-beam codebook $\mathbf{C}$ to combine into a multi-arm beam\cite{Wang7915742}. we denote the universe of keys with total order as $\mathcal{U}=\{0,1,...,N_C-1\}$, the hash function as $h:\mathcal{U}\to\mathcal{T}$, where $\mathcal{T}=\{0,1,...,B-1\}$ is the interpreted hash value, $N_C$ is the number of codewords and $B$ is the number of hash value. Here, $\mathcal{U}$ is fixed and we randomly select hash functions from a family $\mathcal{H}=\{h_1,h_2,...,h_{|\mathcal{H}|}\}$, where $|\mathcal{H}|$ represents the number of distinct hash functions in the family $\mathcal{H}$.
\begin{figure}[t]
	\centering 
	\includegraphics[width=0.97\linewidth]{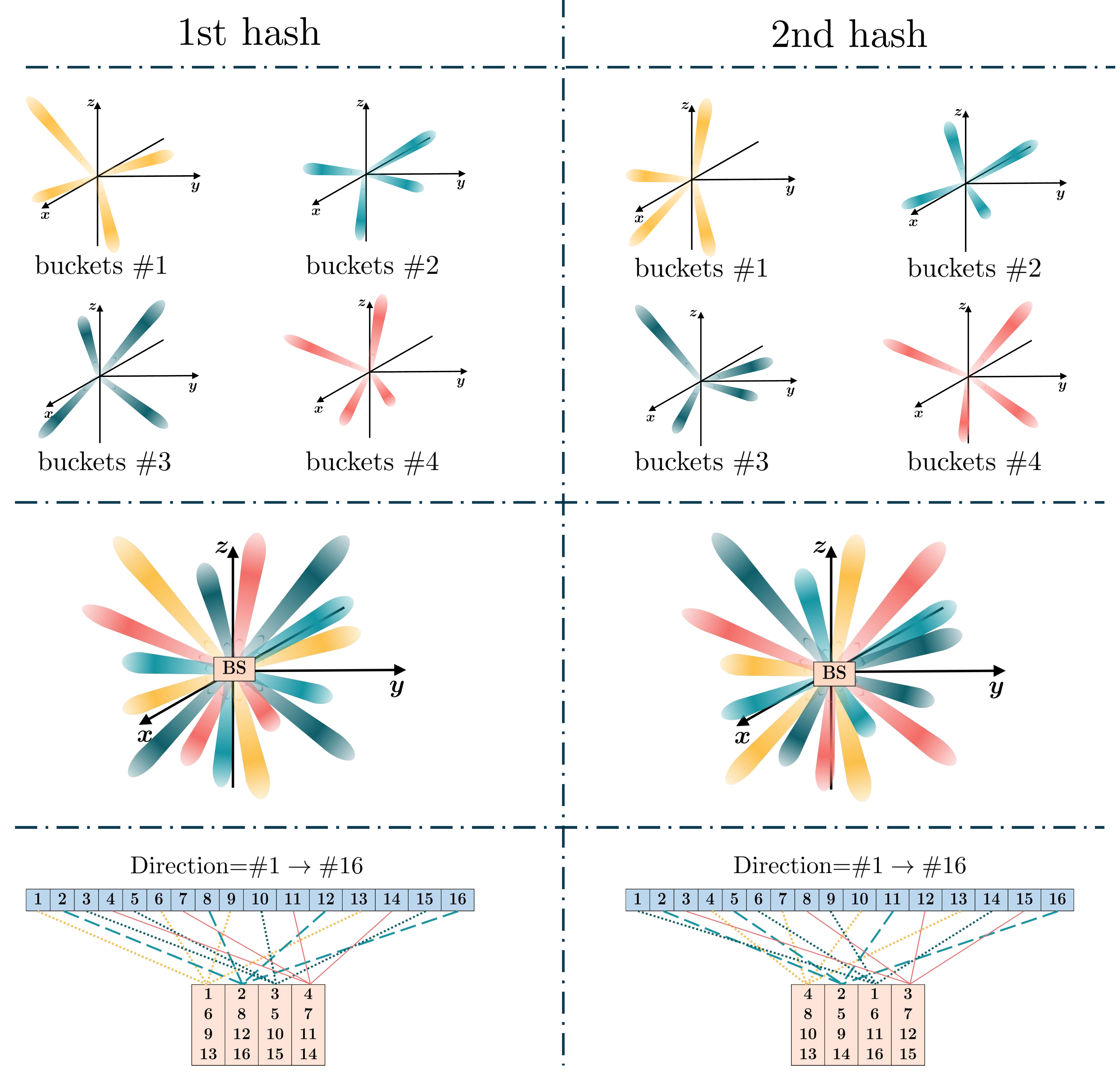}\vspace{-1mm}
	\caption{The schematic diagram for hashing implementation.}\label{fig:hash}\vspace{-2mm}
\end{figure}
\par
Next, we use the hashing results $\mathbf{D}$ to generate the multi-arm beams, where each bucket $\mathbf{d}_b$ containing $R=N_C/B$ keywords corresponding to a hashing multi-arm beam (HMB), can be represented as\vspace{-2mm}
\begin{equation}
	\mathbf{D}=[\mathbf{d}_1;...;\mathbf{d}_B],\quad \mathbf{d}_b=[d_b^1,...,d_b^R],\vspace{-1mm}
\end{equation}
where commas ($,$) and semicolons ($;$) denote the row separators and column separators, respectively.
Fig.~\ref{fig:hash} illustrates two times hash that the codewords representing $16$ directions are hashed uniformly into $4$ multi-arm beams, each covering a total of $N_C/B = 4$ different directions. Specifically, in the first hash, the 16 directions are divided into 4 buckets, which are $[1,6,9,13]$, $[2,8,12,16]$, $[3,5,10,15]$, and $[4,7,11,14]$.

\par
In contrast to the simple antenna partitioning approach\cite{You9129778}, we jointly design the response of all antennas to generate the multi-arm beam codebook $\tilde{\mathbf{C}}$ for training. Specifically, for $\mathbf{d}_b$, we make the digital precoder $\mathbf{f}_{BB_k}$ to map the data stream to the working RF chain, while the analog beamformer $\mathbf{F}_{RF_k}$ selects $V=R$ codewords from the single-beam codebook $\mathbf{C}$ to determine the beams to be transmitted over the working RF chain, i.e., \vspace{-3mm}
\begin{equation}
	[\mathbf{f}^b_{BB_k}]_i=\frac{e^{j\boldsymbol{\vartheta}(i)}}{\sqrt{V}},i=1,...,V,
\end{equation}
\begin{equation}
	\mathbf{F}^b_{RF_k}(:,i)=\mathbf{C}(d_b^i,:)^T,i=1,...,V.\vspace{-1mm}
\end{equation}
\begin{figure*}[t]
\centering\vspace{-4mm}
\begin{equation}\label{eq:y_K} 
	\begin{split}
		|y_k|^2(\mathbf{d}_b,\boldsymbol{\vartheta})&\overset{(a)}{\approx}|\sqrt{P_0}|\beta_1|\sqrt{MN}e^{-j\frac{2\pi}{\lambda_c}r_k}\sqrt{W(r_k,\theta_k,\phi_k,\boldsymbol{\vartheta},\mathbf{d}_b)}  +n|^2\vspace{-4mm}
	\end{split}
\end{equation}
\begin{equation}\label{eq:W}
	W(r_k,\theta_k,\phi_k,\boldsymbol{\vartheta},\mathbf{d}_b)=\sum\limits_{i=1}^{V}\frac{1}{V} e^{j\boldsymbol{\vartheta}(i)}\sum\limits_{n=-\frac{N-1}{2}}^{\frac{N-1}{2}}\sum\limits_{m=-\frac{M-1}{2}}^{\frac{N-1}{2}} \frac{1}{MN} e^{-j\frac{2\pi}{\lambda_c}(D^{d_b^i}(m,n)+D(m,n)-r_{d_b^i}-r_k)}\vspace{-1mm}
\end{equation}
\begin{equation}\label{eq:W'}
	W'(r_k,\theta_k,\phi_k,s)=\sum\limits_{n=-\frac{N-1}{2}}^{\frac{N-1}{2}}\sum\limits_{m=-\frac{M-1}{2}}^{\frac{N-1}{2}} \frac{1}{MN} e^{-j\frac{2\pi}{\lambda_c}(D^{
			s}(m,n)+D(m,n)-r_{d_b^i}-r_k)}\vspace{-4mm}
\end{equation}\vspace{-3mm}
{\noindent} \rule[-10pt]{17.5cm}{0.05em}\\
\end{figure*}
Therefore, the user's received power from BS $k$ can be derived as (\ref{eq:y_K}), where (\ref{eq:W}) is the normalized multi-arm beam radiation pattern of $\mathbf{d}_b$, and (\ref{eq:W'}) is the normalized radiation pattern of the $s$-th single beam.
\par
Because the different near-field single beams interfere with each other in their respective main lobes, the $s$-th near-field single beam has a main lobe region of $\cos\phi'\in[\cos\phi_s-\frac{1}{M},\cos\phi_s+\frac{1}{M}]$, $\cos\theta'\in[\cos\theta_s-\frac{1}{N},\cos\theta_s+\frac{1}{N}]$,
\begin{equation}
    r'\in[\frac{1+\frac{r_s\lambda_c\zeta_\Delta^2}{\sin^2\phi_sM^2d_z^2}}{\frac{1}{r_s}+\frac{2\lambda_c\zeta_\Delta^2}{\sin^2\phi_sM^2d_z^2}},\frac{1-\frac{r_s\lambda_c\zeta_\Delta^2}{\sin^2\phi_sM^2d_z^2}}{\frac{1}{r_s}-\frac{2\lambda_c\zeta_\Delta^2}{\sin^2\phi_sM^2d_z^2}}].
\end{equation}
We define the unit deviation of $W(r_k,\theta_k,\phi_k,\boldsymbol{\vartheta},\mathbf{d}_b)$ and $W'(r_k,\theta_k,\phi_k,d_b^i)$ as\vspace{-1mm}
\begin{equation}\label{eq:delta}
\begin{split}
    &\delta_W(\boldsymbol{\vartheta},d_b^i)\triangleq \int_{r'}\int_{\theta'}\int_{\phi'}\\
    &\ \ |\frac{W(r_k,\theta_k,\phi_k,\boldsymbol{\vartheta},\mathbf{d}_b)-W'(r_k,\theta_k,\phi_k,d_b^i)}{W'(r_k,\theta_k,\phi_k,d_b^i)}|\mathrm{d} \phi\mathrm{d} \theta\mathrm{d}r.
\end{split}
\end{equation}
Further, define the average deviation of the multi-arm beams and single-beam radiation patterns in the corresponding main lobe as\vspace{-2mm}
\begin{equation}
	\delta_W(\boldsymbol{\vartheta},\mathbf{d}_b)\triangleq \frac{1}{V}\sum\limits_{i=1}^{V} \delta(\boldsymbol{\vartheta},d_b^i).
\end{equation}
Therefore, based on the above HMB combination $\mathbf{d}_b$, we can minimize $\delta_W(\boldsymbol{\vartheta},\mathbf{d}_b)$ by adjusting $\boldsymbol{\vartheta}=\boldsymbol{\vartheta}_b$ to produce a well-shaped multi-arm beam, denoted as $\tilde{\mathbf{C}}(b,:)=\mathbf{F}^b_{RF_k}\mathbf{f}^b_{BB_k}$.
\vspace{-1mm}
\section{Near-Field Beam Training}
\vspace{2.5mm}
The training process includes the scanning phase and the voting phase. To ensure the accuracy of beam training, we conduct a total of $L$ rounds of hashing mapping. Specifically, BS $k$ randomly selects $L$ different hash functions $h_1^k,...h_L^k$ from the family $\mathcal{H}$ to obtain $\mathbf{D}^k_1,...,\mathbf{D}^k_L$, and simultaneously sends training symbols with its own multi-arm beam codebook $\tilde{\mathbf{C}}^k_1,...,\tilde{\mathbf{C}}^k_L$ during the scanning phase, significantly reducing the complexity of traditional alternate scanning of all BSs. Thus, a total of $Q=BL$ time slots are needed, which yielding $Q$ received signal power, denoted as $\mathbf{P}=[P(1,1),...,P(l,b),...,P(L,B)]$, where the measurement $P(l,b)$ of the $q=(l-1)B+b$-th time slot is the recorded power of the signal received by the $b$-th multi-arm beam of the $l$-th round of hashing,\vspace{-2mm}
\begin{equation}
	P(l,b)=|\sum\limits_{k=1}^K\mathbf{h}_k^H\mathbf{F}_{RF_k}\mathbf{f}_{BB_k}x+n|^2.\vspace{-2mm}
\end{equation}
In the following, for multi-BS superimposed signals, we design a demultiplexing algorithm to isolate signals from different BSs and use soft decision and the voting mechanism to obtain aligned beams for each BS.
\par
Suppose that the direction of the user with respect to BS $k$ is  $\gamma_k\in\mathcal{U}$, the probability of two arbitrary BSs seeing this user at the same time is $Pr(h_i(\gamma_i)\underset{(i\neq j)}{=}h_j(\gamma_j))=\frac{1}{B^2}$, which is small enough so that the received signal of each time slot almost contains the signal of at most one BS. Also, the different distances from different BSs to this user result in different channel gains, with the same transmit power, we can obtain distinguishable received signal strengths as $\dot{P}_{m_1}>\dot{P}_{m_2}>...>\dot{P}_{m_K}$, where $m_k$ is the BS with the $k$-th strongest channel gain. Based on this, the demultiplexing algorithm can be designed in conjunction with the soft decision, i.e., assigning the $L$ time slot with the $(k-1)L\!+\!1$-th - $kL$-th largest value in the received signal power $\mathbf{P}$ to BS $m_k$, which means\vspace{-2mm}
\begin{equation}
		\mathbf{q}_{m_k}=\arg\max\limits_{(k-1)L+1:kL} descend(\mathbf{P}),
	\vspace{-2mm}
\end{equation} 
where $descend(\cdot)$ represents sorting the vector in descending order. The reason for $L$ is that each BS sees the user in only one time slot per round of hashing. 
\par
Now that we distinguish the received signal power of different BS, we can then conduct voting on $\tilde{\mathbf{D}}^k(\mathbf{q}_{m_k},:)$ to find $\gamma_k$, where colon ($:$) denotes all the elements of the row/column. The reason that the vote can end up with a unique direction is that if we fix $x_1$ and $x_2$ that satisfy $h_i(x_1)=h_i(x_2)$, we have $Pr[h_{i'}(x_1) = h_{i'}(x_2)] = \frac{1}{B}$, where ${i'\neq i}$, $h_i,h_{i'}\in\mathcal{H}$. That is, the probability of arbitrary two keywords $x_1$ and $x_2$ being hashed to the same address simultaneously by different hash functions is sufficiently small. It ensures that multiple rounds of hashing make the directions dispersed from each other. However, based on the demultiplexed time slot $\mathbf{q}_{m_k}$ which contains the aligned beam, it's most likely the direction $\gamma_{m_k}$ that gets the highest votes.
\par
In general, the detailed steps of beam training for multiple users are discussed in Algorithm \ref{alg:Framwork}. Firstly, in the scanning phase, all BSs simultaneously send training symbols utilizing the predefined multi-arm beams with the same power $P_0$, until all the predefined multi-arm beams are traversed. Meanwhile, all users receive from the channel omnidirectionally. Next, we demultiplex the multi-BS superimposed signal power $\mathbf{P}_u$ received by user $u$, by soft decision that assigns the $L$ time slot with the $(k-1)L\!+\!1$-th - $kL$-th largest value in $\mathbf{P}_u$ to BS $m_k$. Afterward, voting on $\tilde{\mathbf{D}}^k(\mathbf{q}_{m_k},:)$ and obtain the highest votes as $\gamma^u_k$, and  the aligned beam of BS $k$ corresponding to user $u$ is the $\gamma^u_k$-th code word in the single beam codebook $\mathbf{C}$, denoted as $\mathbf{C}(\gamma_k^u,:)$.
\vspace{1mm}

\begin{algorithm}
	\caption{HMB Training}
	\label{alg:Framwork}
	\begin{algorithmic}[1]
		\Require 
		\Statex Hashing results for all BSs $\{\mathbf{D}^k_1,...,\mathbf{D}^k_L\}_{k=1}^K$
  \Statex Multi-arm beam codebooks $\{\tilde{\mathbf{C}}_1^k, \tilde{\mathbf{C}}_2^k, \ldots, \tilde{\mathbf{C}}_L^k\}_{k=1}^K$
		\Statex Transmit signal $x$
		\Statex Number of BS $K$
		\Statex Number of hashing rounds $L$ and time slot $Q$
		\Ensure 
		\Statex Aligned beam index $\{\bm{\gamma}^u\}_{u=1}^U$, $\bm{\gamma}^u=[\gamma^u_1,...,\gamma^u_K]$
		\Statex Aligned beam of BSs corresponding to users
		
		\For{$q$ = 1 to $Q$}
		\State $\forall$ BS $k$ transmit $x$ by the $q$-th multi-arm beam in 
  $\{\tilde{\mathbf{C}}_1^k, \tilde{\mathbf{C}}_2^k, \ldots, \tilde{\mathbf{C}}_L^k\}$
		\State all users record the multi-BS superimposed received signal powers $\{\mathbf{P}_u\}_{u=1}^U$
		\EndFor
		\For{($\forall$ user $u$) $k$ = 1 to $K$}
		\State $\mathbf{q}_{m_k}=\arg\max\limits_{(k-1)L+1:kL} descend(\mathbf{P}_u)$
		\State $\gamma^u_k\leftarrow$ most votes on $\tilde{\mathbf{D}}^k(\mathbf{q}_{m_k},:)$
		\State aligned beam of BS $k$ to user $u$ is $\mathbf{C}(\gamma_k^u,:)$
		\EndFor
	\end{algorithmic}
\end{algorithm}
\vspace{-4mm}
\section{Simulation Results}
\vspace{2mm}
We now evaluate the performance of our proposed beam training method with simulation results. The number of BSs and the operating frequency are set to $K=5$ and $f_c=28\text{GHz} $ respectively, and the signal wavelength is $\lambda_c=0.01\text{m}$. The planar antenna array of BS contains $M=4$, $N=128$ antennas, and the spacing between the antennas is $d_x=d_z=\lambda_c/2 $. The reference signal to noise ratio (SNR) is $\gamma=\frac{P_0MN\rho_0}{r_0^2\sigma^2}$ with $\rho_0=-72$dB, $P_0=15$dBm, $\sigma^2=-70
$dBm. And the achievable rate in bits/second/Hz (bps/Hz) is given by $R=log_2(1+\gamma|\mathbf{f}_{BB}^T\mathbf{F}_{RF}^T\mathbf{g}_1|^2)$.
\par
\begin{figure}[t]
	\begin{center}
			\centerline{\includegraphics[width=0.41\textwidth]{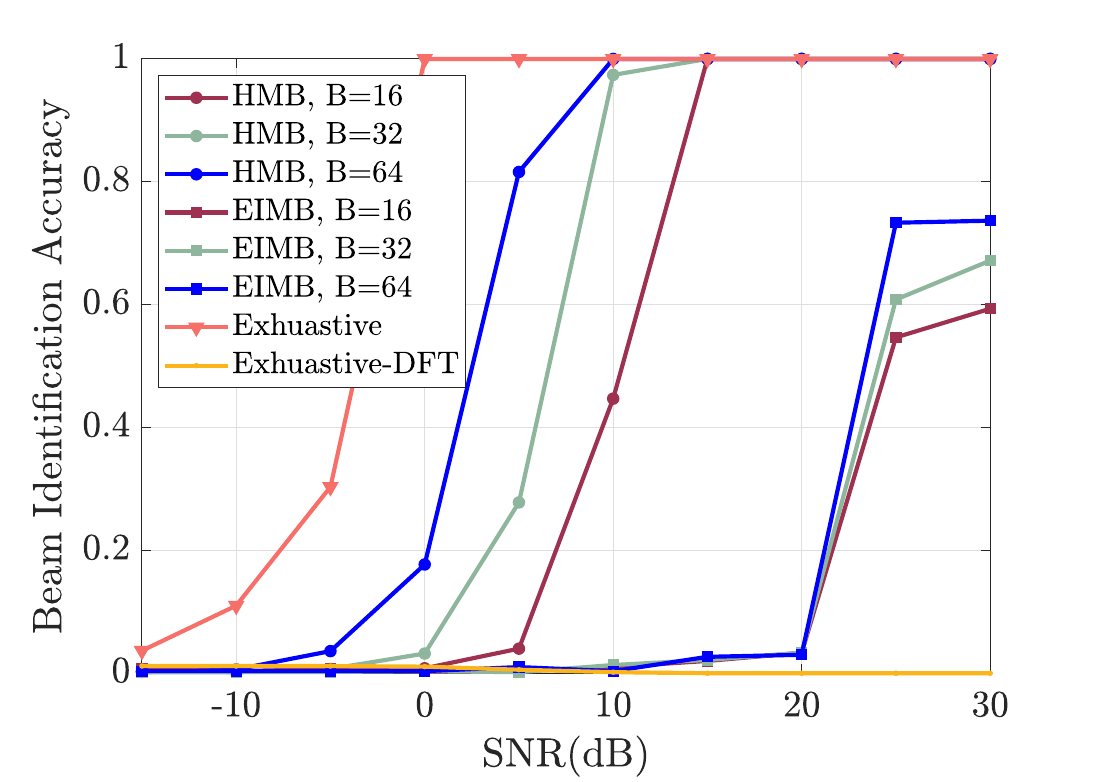}} \vspace{-2mm}
			\caption{Success beam identification accuracy versus SNR.  }
			\label{fig:accuray} \vspace{-10mm}
		\end{center}
\end{figure}

Fig.~\ref{fig:accuray} plots the effect of SNR on the identification accuracy. With the same simulation setup, we use exhaustive, EIMB training with the near-field codebook and exhaustive training with the DFT codebook ("Exhaustive-DFT") as the baseline. Firstly, it can be seen that with increasing SNR, the influence of noise becomes smaller and the identification accuracy of all beam training methods gradually increases; under the exhaustive beam training, the accuracy with the near-field codebook converges to 1, while that with the far-field codebook is significantly lower, which confirms the effectiveness of the designed codebook in near-field conditions.
\par
In addition, the performance of EIMB is quite lower compared to the exhaustive and HMB approaches when utilizing near-field codebooks. In HMB training, when the number of multi-arm beams $B\geq 32$ and the SNR is not less than 5 dB, at least 96.4\% of the performance of exhaustive training can be achieved. Moreover, when the SNR is relatively low, the identification accuracy is considerably improved compared to EIMB training. However, it can be noticed that as the number of multi-arm beams $B$ decreases, the accuracy gradually decreases, the reason being that the number of sub-beams $R$ increases as $B$ decreases, which makes the leakage interference between sub-beams have a greater effect on the identification result.
\begin{figure}[t]
	\begin{center}
			\centerline{\includegraphics[width=0.41\textwidth]{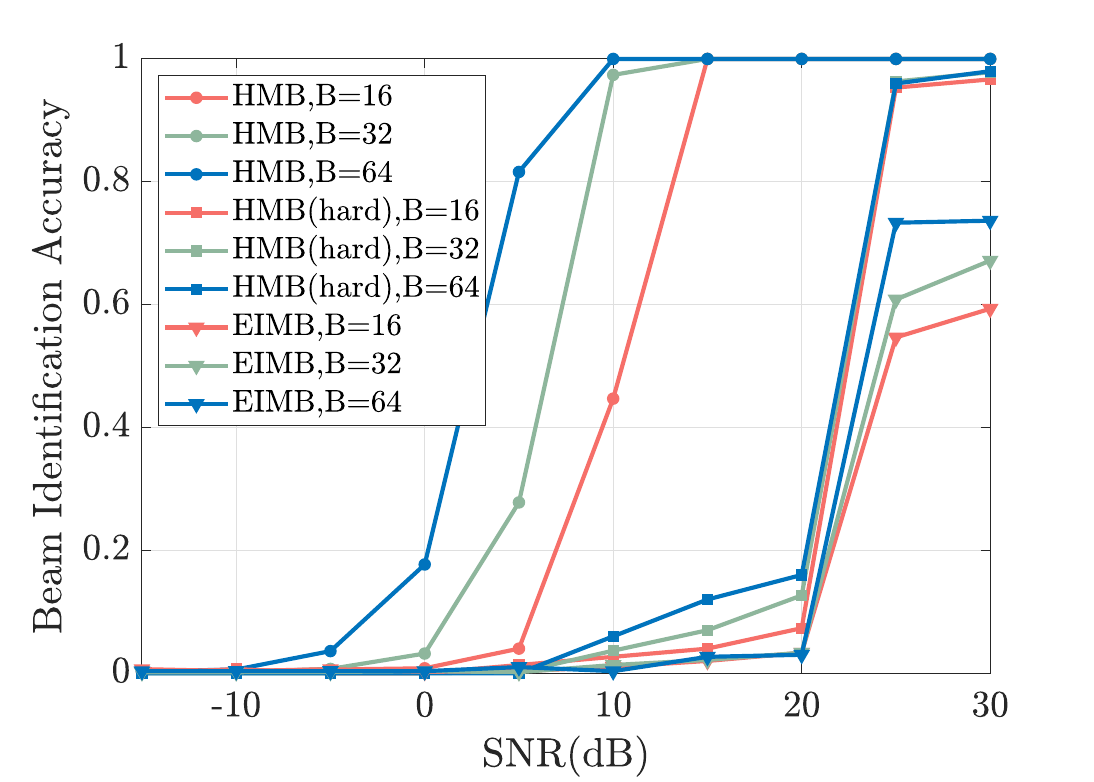}}  \vspace{-2mm}
			\caption{Success beam identification accuracy versus SNR when considering soft and hard decisions.  }\vspace{-8mm}
			\label{fig:accuracy_softhard} 
		\end{center}
\end{figure} 
\par
Fig.~\ref{fig:accuracy_softhard} plots the beam identification accuracy versus SNR when considering soft and hard decisions. where only "HMB" training uses the soft decision, and the others use the hard decision of threshold comparison. It can be seen that our proposed HMB training method has the optimal performance, especially when the SNR is relatively small. Specifically, when the SNR = 10dB and the number of beams $B = 32$, the soft decision can improve the accuracy by 96.9\%. This is because when the SNR is very low, the intensity of the signal power can be of the same order as the noise power, or even submerged in the noise. Thus the threshold of hard decisions needs to be determined more accurately and adaptively, but it is difficult. on the contrary, the soft decision is based on relative value comparisons, which does not need to determine the threshold and is less affected by noise as well. In addition, when the SNR is higher than 20dB, the HMB codebook can improve the accuracy by 22\% over the basis of EIMB codebook, because the equal interval method has a fixed leakage interference, while the randomness of hashing adds a random perturbation to the leakage interference between sub-beams, so that the effect of this interference on the subsequent decision can be reduced. 

\begin{figure}[t]
	\begin{center}
			\centerline{\includegraphics[width=0.41\textwidth]{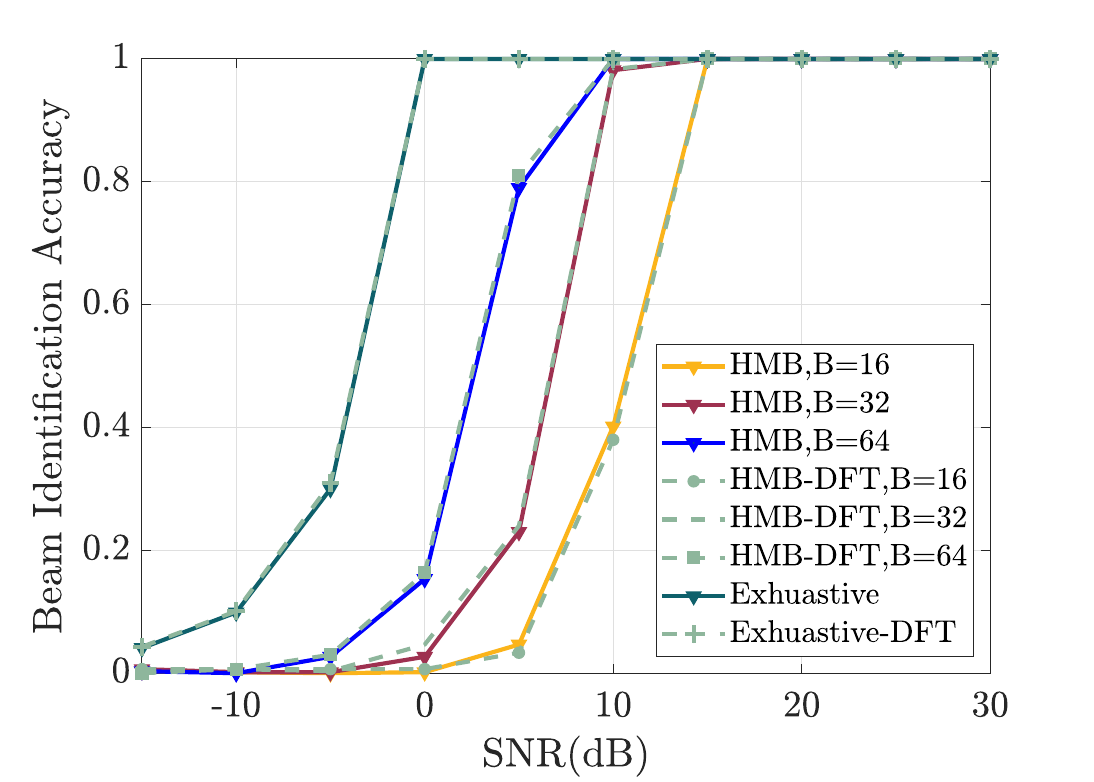}}  \vspace{-2mm}
			\caption{Success beam identification accuracy versus SNR under the far-field simulation condition.}\vspace{-9mm}
			\label{fig:accuracy_far} 
		\end{center}
\end{figure} 
\par
Fig.~\ref{fig:accuracy_far} plots the beam identification accuracy versus SNR under the far-field simulation condition. The Rayleigh distance can be calculated as $Z=81.92$m, therefore, we take the distance between the user and the antenna as $r=300$m. It can be seen that the codebook constructed by our method has almost the same accuracy as the DFT codebook, which verifies the excellent applicability of the proposed HMB training method even in the far-field region.
\begin{figure}[t]
	\begin{center}
			\centerline{\includegraphics[width=0.41\textwidth]{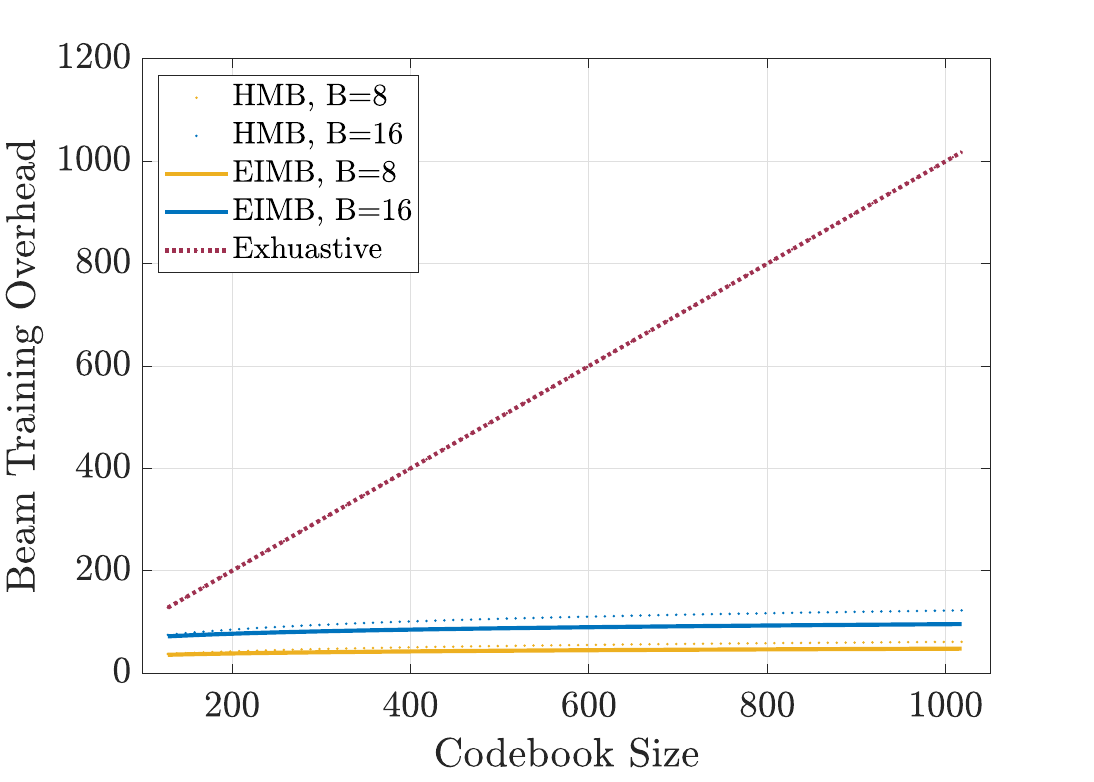}}  \vspace{-2mm}
			\caption{Training overhead versus codebook size.  }\vspace{-8mm}
			\label{fig:overhead} 
		\end{center}
\end{figure} 
\par
Fig.~\ref{fig:overhead} plots the effect of different codebook sizes on the beam training overhead, where the training overhead is defined as the number of time slots required for scanning during training. The exhaustive approach traverses the entire beam space and the training time is proportional to the codebook size, resulting in a very high training overhead. HMB training is at the logarithmic level with a training time of $Q=BL=O(B\mathrm{log}MN)$, which can significantly reduce the training overhead compared to exhaustive beam training. Although the training overhead of EIMB is low in the figure, it presents a much lower accuracy.
\vspace{-1mm}
\section{Conclusion}
\vspace{2.mm}
In this paper, the HMB training method was proposed for the near-field and verified to be applicable to the far-field as well. Firstly, by exploiting the polar domain sparse property of the near-field steering vectors, we minimized the projection between the vectors at different sampling points and constructed the training beams for the near field. To further improve the performance of beam training, we use hash functions to generate multi-arm beams and employ the soft decision and voting mechanism to obtain the best-aligned codeword to maximize the received SNR. Simulation results show that our proposed beam training method has maintained stable and satisfactory performance in terms of beam identification accuracy, reaching 96.4\% out of the exhaustive training performance while ensuring that the training overhead is significantly reduced to the logarithmic level.

\bibliographystyle{IEEEbib}
\bibliography{myrefs}
\end{document}